\documentclass[reqno, a4paper]{amsart}

\usepackage{enumerate}
\usepackage{natbib}
\usepackage[margin=2.5cm]{geometry}
\usepackage{booktabs}
  \usepackage{amssymb,mathrsfs,nicefrac}
  \providecommand{\R}{\mathbb{R}}
  \providecommand{\ccF}{\mathcal{F}}
  \providecommand{\Ind}{\mathbf{1}}
  \DeclareMathOperator{\Var}{Var}

\usepackage{xcolor}
\usepackage[colorlinks,urlcolor=red,citecolor=blue,linkcolor=red]{hyperref}
\hypersetup{hidelinks}

\title{Arbitrage in Estimate Nothing: an example}
\author{Johannes Brutsche$^{1}$ \and  Thorsten Schmidt$^{1}$ \and Julian Sester$^{2}$}
\date{\today}

\thanks{The counterexample was constructed with the aid of ChatGPT}

\usepackage{todonotes}

\begin{document}

\maketitle

\begin{center}
$^{1}$Department of Mathematical Stochastics, Mathematical Institute,\\
University of Freiburg, Ernst-Zermelo-Str. 1, 79104 Freiburg, Germany.\\[2mm]
$^{2}$National University of Singapore, Department of Mathematics,\\ 21 Lower Kent Ridge Road, 119077, Singapore                                                         \end{center}

\begin{abstract}
We give a two-period counterexample to the absence of arbitrage for the
posterior-weighted pricing rule in \emph{Estimate nothing} by Duembgen and
Rogers. Both physical models have strictly positive transition densities,
and each model is equipped with an equivalent martingale measure. Nevertheless,
the mixed price of a single derivative falls deterministically from $5/2$
to $2$ between two trading dates. If these prices are tradable, shorting
the derivative and closing the position one period later yields a certain
profit. A finite-state appendix also illustrates the failure of recursive
consistency.
\end{abstract}

\section{Introduction}
The work of \cite{DuembgenRogers2014} received great attention over time. It proposes a mixture procedure
from different models to obtain pricing and hedging routines. In this note,
we examine whether the resulting prices are necessarily free of dynamic
arbitrage. A concrete two-period example shows that this need not hold,
even when all component models are arbitrage-free and have strictly positive
transition densities. The conclusion concerns the interpretation of the
output as a tradable price process; for non-tradable valuation estimates,
the example instead shows a failure of dynamic pricing consistency.

\section{The framework of Duembgen and Rogers}

Consider a discrete-time financial market with zero interest rates and a single non-dividend-paying asset $S$ whose log-price process we denote by $X=(X_t)_{t=0,h,2h,\dots}$. 

There are $J$ possible candidate {physical models under which $X$ is a time-homogeneous Markov process}, whose {transition} densities are denoted by
\begin{align}
	p_j(x,x') = \frac{P_j(X_h \in dx' | X_0 = x)}{dx'}, \qquad j=1,\dots,J. 
\end{align}
For models with stochastic volatility, $X$ might even be modelled higher-dimensional. Associated to model $j$ is a pricing function which gives {the price} $\varphi_j^a(x)$ {of} derivative $a \in \{1,\dots,A\}$ given the current log-price $x$.

The {log-likelihood score of model $j$} is defined as
\begin{align}\label{eq:log-likelihood}
	\ell^j_t &= \beta \ell^{j}_{t-h} + \log p_j(X_{t-h},X_t) - \tilde Q(\varphi_j(X_t),Y_t),
\end{align}
with $\ell^j_0=0$ and $\tilde Q$ being a non-negative quadratic form, $Y_t^a$ the market price of derivative $a$ and $\beta \in (0,1]$. 

{The model weights are given by the normalised likelihood scores below.
For $\beta=1$ and $\tilde Q\equiv0$, they are the usual Bayesian posterior
probabilities based on the observed asset prices. For $\beta<1$, the recursion
instead discounts past observations:}
\begin{align}
	\pi^j_t &= \frac{\exp(\ell^j_t)}{\sum_{i=1}^J \exp(\ell^i_t)}, \qquad j=1,\dots,J,\ {t\in\{0,h,2h,\dots\}}. 
\end{align}
In particular, $\pi^j_0=J^{-1}$ for all $j=1,\dots,J$, i.e.\ the prior at time $0$ is {uniform}.

\cite{DuembgenRogers2014} propose to price an exotic derivative with payoff $H$ by calculating price processes $V^j(H)$ under model $j$ {and then taking the weighted average. Here we assume
$V_t^j(H)=E_{Q_j}[H\mid\ccF_t]$, where $\ccF_t$ is the observed filtration,
$Q_j$ is an equivalent martingale measure for model $j$, and the payoff $H$
at maturity $T$ is integrable under every $Q_j$. The proposed rule is}
\begin{align}
	\label{eq:DR pricing rule}
	V_t(H) \equiv \sum_{j=1}^J \pi^j_t V^j_t(H), \qquad {t\in\{0,h,2h,\dots,T\}}.
\end{align}
{Below}, in Section~\ref{eq:counterexample}, we show that this construction {need not be arbitrage-free if the derivative can be traded at these prices}.

\subsection{The counterexample}\label{eq:counterexample}
We now specialise to two periods, two models and choose $h=1$.
{We work on $\Omega=(0,\infty)^2$, with coordinate prices $(S_1,S_2)$,
$\ccF_t=\sigma(S_1,\ldots,S_t)$ for $t=1,2$, and trivial $\ccF_0$.
The bank account is $B_t=1$. Trading takes place at $0,1,2$, without
transaction costs or short-sale constraints. We take $\beta=1$; since
$\ell_0^j=0$, the time-one calculation is unchanged for any $\beta\in(0,1]$.} Moreover, we assume that the calibration quality of the two models {coincides}, i.e.\footnote{
Under the martingale measures chosen later, see \eqref{ex:Q}, this typically will only hold for selected derivatives. {In particular, choosing $\tilde Q\equiv0$ makes this equality automatic.}
} 
$$ \tilde Q(\varphi_1(X_1),Y_1)=\tilde Q(\varphi_2(X_1),Y_1)=:\tilde Q_1. $$
Then 
\begin{align}\label{eq_pi1}
	{\pi_1^1} &=  \frac{p_1(X_0,X_1)e^{-{\tilde Q_1}}}{ (p_1(X_0,X_1)+p_2(X_0,X_1))e^{-{\tilde Q_1}}} 
	= \frac{p_1(X_0,X_1)}{p_1(X_0,X_1)+p_2(X_0,X_1)}.
\end{align}

The asset price starts with $S_0=2$. 
We directly assume
$$ p_1(x,x')=e^{2x'-e^{x'}}, \qquad p_2(x,x')=e^{x'-e^{x'}}, $$
which are independent of the prior state $x$. {The substitution $s'=e^{x'}$ gives the conditional densities of $S_t$,
\[
 f_1(s')=s'e^{-s'},\qquad f_2(s')=e^{-s'},\qquad s'>0.
\]
Their integrals are both one. Thus $p_1$ and $p_2$ are strictly positive
densities on $\R$; under $P_1$ the prices $S_1,S_2$ are independent
$\operatorname{Gamma}(2,1)$ variables (shape $2$, rate $1$), and under $P_2$
they are independent exponential variables of rate $1$.} 

For the moment note that by \eqref{eq_pi1},
\begin{align}\label{eq:computation_pi}
	{\pi_1^1}=\frac{ e^{2x'} \cdot e^{-e^{x'}}}
	{( e^{2x'} + e^{x'}) \cdot e^{-e^{x'}}} = \frac{e^{x'}}{e^{x'}+1}=\frac{s}{1+s},
\end{align}
{with $x'=X_1$ and $s=S_1=e^{X_1}$. Similarly, $\pi_1^2=(1+s)^{-1}$.} 

Next, we choose the martingale measures for each model. We consider the returns
$$ R_t = \frac{S_{t}}{S_{t-1}} $$
{and define $Q_j$ by Markov transition kernels such that, at each
$t\in\{1,2\}$, the conditional distribution of $R_t$ given $\ccF_{t-1}$
depends only on $S_{t-1}=s$ and is log-normal, i.e.}
\begin{align}\label{ex:Q}
	\log R_t \sim_{Q_j|S_{t-1}=s}  \mathscr{N}(-\nicefrac 1 2\,  \sigma_j^2(s), \sigma_j^2(s)).
\end{align}
Then $E_{Q_j}[R_t|S_{t-1}]=1$ and hence each $Q_j$ is a martingale measure. {Moreover, each $Q_j$ is equivalent to each $P_i$ on $\ccF_2$:
the one-step price densities under $Q_j$ are strictly positive on
$(0,\infty)$, as are $f_1$ and $f_2$. Consequently, all four measures
have strictly positive joint densities on $(0,\infty)^2$ with respect
to Lebesgue measure.} 
Moreover, we choose ${\sigma_1^2(s)}=\log 2$ and $\sigma_2^2(s)=\log(3+s)$.

To construct an arbitrage, we consider a  derivative with path-dependent payoff 
$$ H: =  \bigg(\frac{S_2-S_1}{S_1} \bigg)^2 =(R_2-1)^2.$$
Since $\Var_{Q_j}(R_2|S_{1}=s)=e^{\sigma_j^2(s)}-1, $
\begin{align}
	V_1^1(H)=1, \qquad V_1^2(H) = 2+S_1.
\end{align}
Hence, since $S_0 =2$, and by using $V_0^2(H)=E_{Q_2}[V_1^2(H)]$, its prices at time $0$ are
\begin{align}
	V_0^1(H)=1, \qquad V_0^2(H)=2+2 = 4.
\end{align}
{These finite expectations also verify that $H\in L^1(Q_1)\cap L^1(Q_2)$.}
This implies the Duembgen-Rogers price
\begin{align}
	V_0(H)=\frac{1}{2} V_0^1(H) + \frac{1}{2} V_0^2(H)=\frac{5}{2}.
\end{align}
At time $1$, the weights get updated and the Duembgen-Rogers price becomes by \eqref{eq:computation_pi}
\begin{align}
	V_1(H) = \pi_1^1 V_1^1(H) + \pi_1^2 V_1^2(H) = 
	\frac{S_1}{1+S_1} \cdot 1 + \frac{1}{1+S_1} \cdot (2 + S_1) = 2.
\end{align}
{This yields an explicit self-financing arbitrage. At time $0$, short one
unit of the derivative for $5/2$ and place the proceeds in the bank account,
so that the initial portfolio value is zero. At time $1$, repurchase the
derivative for its deterministic price $2$. The remaining bank balance is
\[
 \frac52-2=\frac12,
\]
which can be held until time $2$. The strategy has non-negative wealth at
every trading date and a strictly positive terminal payoff on every path.
It requires no trading in the underlying and no knowledge of which physical
model is correct.

The failure results from applying weights updated using physical
likelihoods to risk-neutral conditional valuations. Pricing instead by
$E_{\bar Q}[H\mid\ccF_t]$ under a fixed mixture
$\bar Q=\tfrac12Q_1+\tfrac12Q_2$ would be dynamically consistent and
arbitrage-free, but its conditional model weights generally differ from
the physical posterior weights used above.}

\begin{appendix}
\section{\texorpdfstring{{Finite-state counterexamples}}{Finite-state counterexamples}}

{For completeness, we provide a finite-state example from \cite{Brutsche2018}. It illustrates a failure of recursive consistency,
but does not satisfy the Lebesgue-density assumption in the framework of
\cite{DuembgenRogers2014}.}

\label{ex:trinomial-counterexample}
We consider a two-period trinomial model with trading dates
$0,1,2$. Let
\[
\Omega=\{u,m,d\}^2,
\qquad
u=1.1,\quad m=1,\quad d=0.9,
\]
and define
\[
S_{t}
=
S_0\prod_{k=1}^{{t}} \xi_k,
\qquad
S_0=100,
\qquad
t=0,1,2,
\]
where $\xi_k\in\{u,m,d\}$. The interest rate $r$ is
assumed to be zero and we set $\ccF_{t}=\sigma(\xi_1,\ldots,{\xi_t})$ for $t=1,2$. {We take $\ccF_0$ to be trivial and the empty product to be one.}
Moreover, we specify two physical models {with independent returns $\xi_1,\xi_2$} via their one-step transition probabilities that are given as
\[
\begin{array}{cccc}
 & u & m & d\\ \midrule
p_1 & \frac38 & \frac12 & \frac18\\[1mm]
p_2 & \frac18 & \frac12 & \frac38
\end{array}
\]

For the log-likelihood in~\eqref{eq:log-likelihood}, we take $\beta=1$ and set the quadratic loss term ${\tilde Q}$ equal to zero. Hence,
given prior probabilities $\pi_0(p_1)$ and
$\pi_0(p_2)=1-\pi_0(p_1)$ on these models {(allowing unequal priors by setting $\ell_0^j=\log\pi_0(p_j)$)}, the posterior probability after the first
observation is given by Bayes' formula as
\begin{align}\label{eq:example_posterior}
    \pi_1(p_1) = \frac{p_1(\xi_1)\pi_0(p_1)} {p_1(\xi_1)\pi_0(p_1)+p_2(\xi_1)\pi_0(p_2)},
\end{align}
and $\pi_1(p_2)=1-\pi_1(p_1)$.

Last but not least, we introduce measures $Q_1$ and $Q_2$ as
\[
Q_1(\omega)
=
q_1(\omega_1)q_1(\omega_2),
\qquad
q_1(u)=q_1(d)=\frac14,
\qquad
q_1(m)=\frac12,
\]
and
\[
 Q_2(\omega)
=
q_2(\omega_1)q_2(\omega_2),
\qquad
q_2(u)=q_2(d)=\frac38,
\qquad
q_2(m)=\frac14.
\]
It is easily verified that $E_{Q_j}[\xi_k]=1$ for $j=1,2$ so that the discounted stock-price process is a martingale under both
measures, i.e. both of them are equivalent martingale measures for $p_1$ and $p_2$.
For any bounded $\ccF_{2}$-measurable claim $H$, define the
posterior-weighted valuation by
\begin{equation}
\label{eq:posterior-weighted-valuation}
V_t(H)
:=
\sum_{j=1}^2
\pi_t(p_j)\,
E_{Q_j}
       \bigl[H\mid\ccF_t\bigr],
\qquad
t\in\{0,1\}{,}
\end{equation}
and recursive consistency {of this valuation family} would then require
\begin{align}\label{eq:consisten_valuation}
    V_0\bigl( V_1(H)\bigr) =V_0(H)
\end{align}
for each {bounded} $\ccF_2$-measurable claim $H$. Subsequently, we present two choices of prior probabilities and claims for which condition~\eqref{eq:consisten_valuation} is not met in the previously specified model. In particular, these examples reveal that $(\pi_t(p_1))_{t=0,1}$ being a $Q_1$- and $Q_2$-martingale still allows to violate~\eqref{eq:consisten_valuation}.\smallskip

\begin{enumerate}
\item[(i)]
We choose the prior probabilities as
\[
\pi_0(p_1)=\frac14,
\qquad
\pi_0(p_2)=\frac34,
\]
and by~\eqref{eq:example_posterior} one obtains for
\[
\pi_1(p_1)
=
\begin{cases}
\frac12, & \xi_1=u,\\[1mm]
\frac14, & \xi_1=m,\\[1mm]
\frac1{10}, & \xi_1=d.
\end{cases}
\]
This posterior process is not a martingale under either risk-neutral
measure as
\[
 E_{ Q_1}
   \bigl[\pi_1(p_1)\mid\ccF_0\bigr]
=
\frac12\cdot\frac14
+\frac14\cdot\frac12
+\frac1{10}\cdot\frac14
=
\frac{11}{40}
\neq
\frac14,
\]
and
\[
E_{ Q_2}
   \bigl[\pi_1(p_1)\mid\ccF_0\bigr]
=
\frac12\cdot\frac38
+\frac14\cdot\frac14
+\frac1{10}\cdot\frac38
=
\frac{23}{80}
\neq
\frac14.
\]
Consider the event
\[
A:=\{S_{2}=121\}=\{(\xi_1,\xi_2)=(u,u)\}
\]
together with the respective claim $\Ind_A$. Its time-zero posterior-weighted value is
\begin{align*}
 V_0(\Ind_A)
&=
\frac14\, Q_1(A)
+
\frac34\, Q_2(A) =
\frac14\left(\frac14\right)^2
+
\frac34\left(\frac38\right)^2
=
\frac{31}{256}.
\end{align*}
At time $1$, the event $A$ is still possible only on the branch
$\{\xi_1=u\}$. Since $\pi_1(p_1)=1/2$ on that branch,
\begin{align*}
 V_1(\Ind_A)
&=
\mathbf 1_{\{\xi_1=u\}}
\left(
\frac12\cdot\frac14
+
\frac12\cdot\frac38
\right)=
\frac5{16}\,
\mathbf 1_{\{\xi_1=u\}},
\end{align*}
and it follows that
\begin{align*}
 V_0\bigl( V_1(\Ind_A)\bigr)
&=
\frac14\,
E_{ Q_1}
   \bigl[ {V_1(\Ind_A)}\bigr]
+
\frac34\,
 E_{ Q_2}
   \bigl[ {V_1(\Ind_A)}\bigr]=
\frac5{16}
\left(
\frac14\cdot\frac14
+
\frac34\cdot\frac38
\right)
=
\frac{55}{512}.
\end{align*}
As this differs from the value of $V_0(\Ind_A)$, posterior-weighted valuation does not satisfy the tower property~\eqref{eq:consisten_valuation}. \smallskip

\item[(ii)]
Now suppose that the two models have equal prior probabilities,
\[
\pi_0(p_1)=\pi_0(p_2)=\frac12.
\]
This is, in particular, the case corresponding to
$\ell_0^1=\ell_0^2=0$ in~\eqref{eq:log-likelihood}. Formula
\eqref{eq:example_posterior} gives for this choice
\[
\pi_1(p_1)
=
\begin{cases}
\frac34, & \xi_1=u,\\[1mm]
\frac12, & \xi_1=m,\\[1mm]
\frac14, & \xi_1=d.
\end{cases}
\]
In this case, the posterior probability is a martingale under both
risk-neutral measures:
\begin{align*}
 E_{ Q_1}
   \bigl[\pi_1(p_1)\mid\ccF_0\bigr]
&=
\frac34\cdot\frac14
+
\frac12\cdot\frac12
+
\frac14\cdot\frac14
=
\frac12,\\
 E_{ Q_2}
   \bigl[\pi_1(p_1)\mid\ccF_0\bigr]
&=
\frac34\cdot\frac38
+
\frac12\cdot\frac14
+
\frac14\cdot\frac38
=
\frac12.
\end{align*}
{This time, we consider} the European call option
\[
C=(S_{2}-100)^+.
\]
Its pay-off is non-zero along the paths $(u,u), (u,m),(m,u)$ and therefore the time-zero posterior-weighted {value} is given by
\begin{align*}
 V_0(C)
&=
\frac12\,
 E_{ Q_1}[C]
+
\frac12\,
 E_{ Q_2}[C]\\
&=
21\left[
\frac12\left(\frac14\right)^2
+
\frac12\left(\frac38\right)^2
\right]
+
2\cdot 10\left[
\frac12\cdot\frac14\cdot\frac12
+
\frac12\cdot\frac38\cdot\frac14
\right]=
\frac{553}{128}.
\end{align*}
On the other hand, its posterior-weighted time-one value is
\[
 V_1(C)
=
\begin{cases}
\displaystyle
21\left(
\frac34\cdot\frac14
+
\frac14\cdot\frac38
\right)
+
10\left(
\frac34\cdot\frac12
+
\frac14\cdot\frac14
\right)
=
\frac{329}{32},
& \xi_1=u,\\[5mm]
\displaystyle
10\left(
\frac12\cdot\frac14
+
\frac12\cdot\frac38
\right)
=
\frac{25}{8},
& \xi_1=m,\\[4mm]
0,
& \xi_1=d.
\end{cases}
\]
Applying the time-zero valuation operator to this time-${1}$ value gives
\begin{align*}
 V_0\bigl( {V_1(C)}\bigr)
&=
\frac{329}{32}
\left(
\frac12\cdot\frac14
+
\frac12\cdot\frac38
\right)
+
\frac{25}{8}
\left(
\frac12\cdot\frac12
+
\frac12\cdot\frac14
\right)=
\frac{2245}{512}.
\end{align*}
Again, $V_0\bigl( V_1(C)\bigr) \neq V_0(C)$, violating the consistency condition~\eqref{eq:consisten_valuation}. Thus, even when the posterior probability process is a martingale under
each of the two risk-neutral measures, posterior-weighted valuation need
not be recursively consistent.
\end{enumerate}

{These tower-property violations concern the valuation family across claims.
If both $H$ and the claim with terminal payoff $G=V_1(H)$ can be traded
at $V_t(H)$ and $V_t(G)$, they also yield arbitrage: since $G$ is
$\ccF_1$-measurable, $V_1(G)=G=V_1(H)$. At time $0$, short the more
expensive claim, buy the cheaper one, and invest the difference. Close
both positions at time $1$ at equal prices. The certain gains are $7/512$
in case (i) and $33/512$ in case (ii). This additional tradability
assumption is unnecessary in the continuous-state example, which uses
only one derivative and the bank account.}

\end{appendix}

\end{document}